\documentclass{PoS}

\usepackage{epsfig}
\usepackage{amsmath}
\usepackage{latexsym}
\usepackage{amssymb}
\usepackage{dsfont}
\usepackage{color}
\usepackage{multirow}

\newcommand{\ud}{\mathrm{d}}

\title{Quark phase-space distributions and orbital angular momentum}

\ShortTitle{Quark phase-space distributions and OAM}

\author{\speaker{C\'edric Lorc\'e}\\
        IPNO, Universit\'e Paris-Sud, CNRS/IN2P3, 91406 Orsay, France\\
        and LPT, Universit\'e Paris-Sud, CNRS, 91406 Orsay, France\\
        E-mail: \email{lorce@ipno.in2p3.fr}}

\author{Barbara Pasquini\\
        Dipartimento di Fisica, Universit\`a degli Studi di Pavia, Pavia, Italy\\
        and Istituto Nazionale di Fisica Nucleare, Sezione di Pavia, Pavia, Italy\\
        E-mail: \email{Barbara.Pasquini@pv.infn.it}}

\abstract{We discuss the Wigner functions of the nucleon which provide multi-dimensional images of the quark distributions in phase space. They combine in a single picture all the information contained in the generalized parton distributions (GPDs) and the transverse-momentum dependent parton distributions (TMDs). In particular, we present results for the distribution of unpolarized quarks in a longitudinally polarized nucleon obtained in a light-cone constituent quark model. We show how quark orbital angular momentum can be extracted from these distributions and compare it with alternative definitions given in terms of the GPDs and the TMDs.}

\FullConference{Sixth International Conference on Quarks and Nuclear Physics,\\
		April 16-20, 2012\\
		Ecole Polytechnique, Palaiseau, Paris}

\begin{document}

\section{Introduction}
\label{section-1}

The quantum phase-space or Wigner distributions encode in a unified picture the information obtained from the transverse-momentum dependent parton distributions (TMDs) and the generalized parton distributions (GPDs) in impact-parameter space. The concept of Wigner distributions in QCD was first explored in refs.~\cite{Ji:2003ak} where relativistic effects were neglected. Recently, we identified the impact-parameter representation of the generalized transverse-momentum dependent parton distributions (GTMDs)~\cite{Meissner:2009ww} with the five-dimensional Wigner distributions (two position and three momentum coordinates) which are not plagued by relativistic corrections~\cite{Lorce:2011kd}. Even though the Wigner distributions do not have a strict probablisitic interpretation due to the uncertainty principle, they encode a variety of information and can often be interpreted with semiclassical pictures.

The aim of this contribution is to investigate the phenomenology of the quark Wigner distributions based on successful relativistic quark models~\cite{Lorce:2011dv}, since so far it is not known how to access these distributions directly from experiments. We discuss in particular how the quark orbital angular momentum (OAM) can be extracted from the Wigner distributions, and compare it with alternative definitions based on the GPDs and the TMDs.

\section{Wigner distributions}

We define the quark Wigner distributions $\rho^{[\Gamma]}(\vec b_\perp,\vec k_\perp,x,\vec S)$ in a nucleon with polarization $\vec S$ as the following matrix elements~\cite{Lorce:2011kd}
\begin{equation}\label{wigner}
\rho^{[\Gamma]}(\vec b_\perp,\vec k_\perp,x,\vec S)\equiv\int\frac{\ud^2\Delta_\perp}{(2\pi)^2}\,\langle p^+,\tfrac{\vec\Delta_\perp}{2},\vec S|\widehat W^{[\Gamma]}(\vec b_\perp,\vec k_\perp,x)|p^+,-\tfrac{\vec\Delta_\perp}{2},\vec S\rangle,
\end{equation}
where the Hermitian quark Wigner operator $\widehat W^{[\Gamma]}(\vec b_\perp,\vec k_\perp,x)$ at a fixed light-cone time $y^+=0$ is defined similarly to refs.~\cite{Ji:2003ak}
\begin{equation}\label{wigner-operator}
\widehat W^{[\Gamma]}(\vec b_\perp,\vec k_\perp,x)\equiv\frac{1}{2}\int\frac{\ud z^-\,\ud^2z_\perp}{(2\pi)^3}\,e^{i(xp^+z^--\vec k_\perp\cdot\vec z_\perp)}\,\overline{\psi}(y-\tfrac{z}{2})\Gamma\mathcal W\,\psi(y+\tfrac{z}{2})\big|_{z^+=0},
\end{equation}
with $y^\mu=[0,0,\vec b_\perp]$, $p^+$ the average nucleon longitudinal momentum and $x=k^+/p^+$ the average fraction of nucleon longitudinal momentum carried by the active quark. The superscript $\Gamma$ stands for any twist-two Dirac operator $\Gamma=\gamma^+,\gamma^+\gamma_5,i\sigma^{j+}\gamma_5$ with $j=1,2$. Finally, a Wilson line $\mathcal W\equiv\mathcal W(y-\tfrac{z}{2},y+\tfrac{z}{2}|n)$ ensures the color gauge invariance. As outlined in ref.~\cite{Lorce:2011kd}, such matrix elements can be interpreted as two-dimensional Fourier transforms of the GTMDs in the impact-parameter space. Although the GTMDs are in general complex-valued functions, their two-dimensional Fourier transforms are always real-valued functions, in accordance with their interpretation as phase-space distributions. 

There are in total 16 Wigner functions at twist-two level, corresponding to all the 16 possible configurations of nucleon and quark polarizations. In order to emphasize the link with the quark OAM, we focus on the distortion in the distribution of unpolarized induced by the longitudinal polarization of the proton $\rho_{LU}=\rho^{[\gamma^+]}(\vec b_\perp,\vec k_\perp,x,+\vec e_z)-\rho^{[\gamma^+]}(\vec b_\perp,\vec k_\perp,x,-\vec e_z)$. Such an effect is new since it cannot be accessed at the twist-2 level by the GPDs and the TMDs. Other configurations for the quark and proton polarizations can be found in ref.~\cite{Lorce:2011kd}.
\begin{figure}[th!]
	\centering
		\includegraphics[width=.45\textwidth]{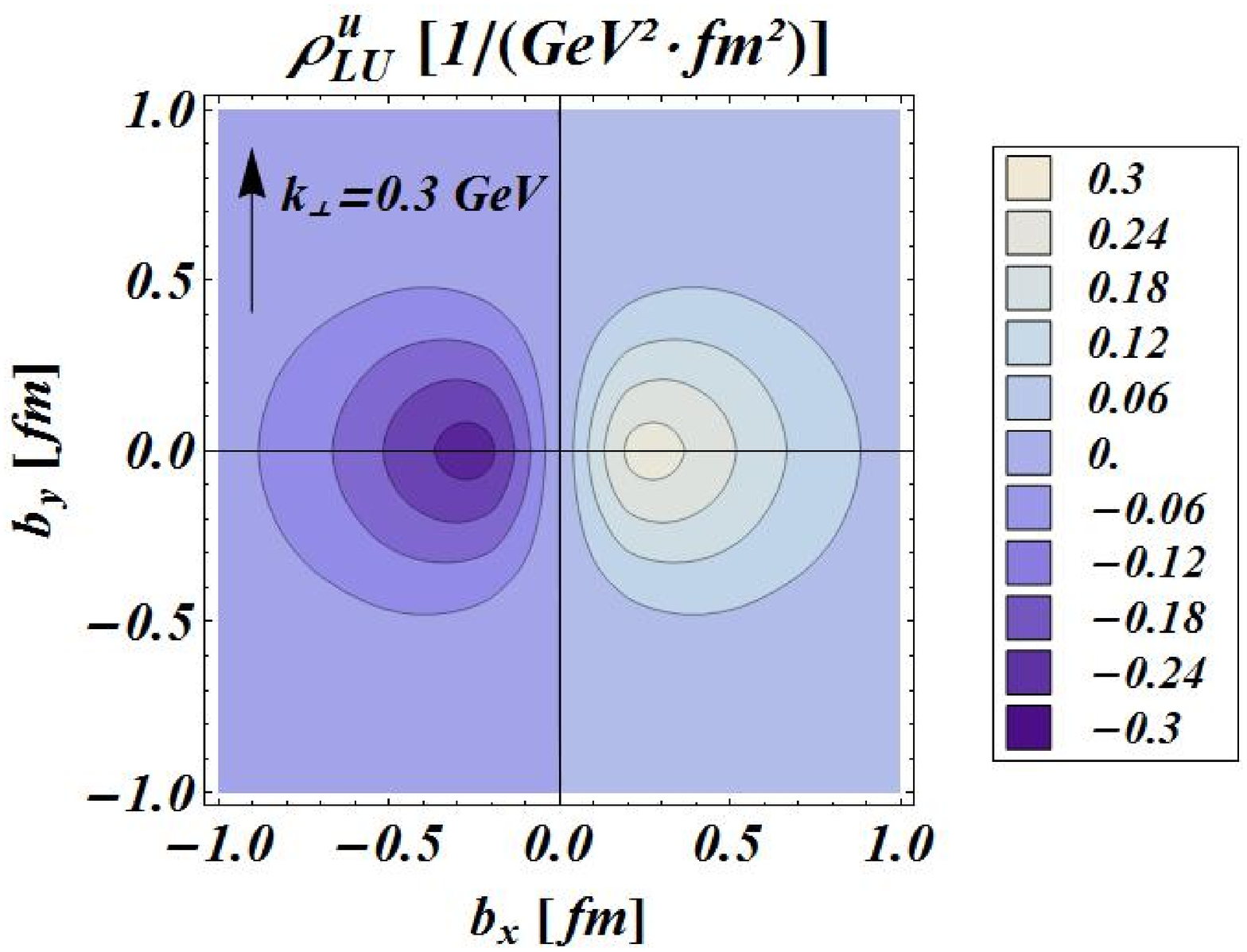}
		\includegraphics[width=.45\textwidth]{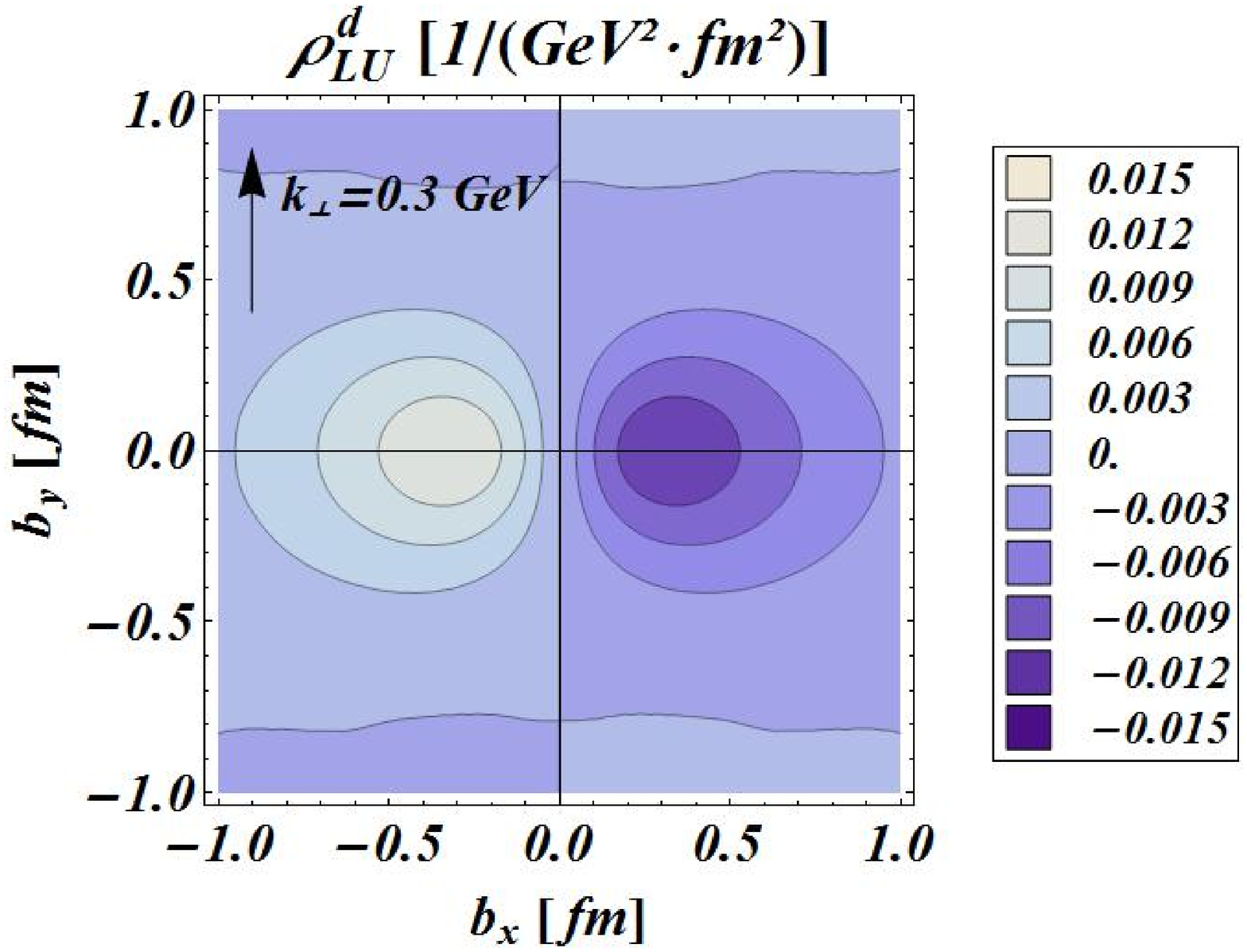}
                     \includegraphics[width=.38\textwidth]{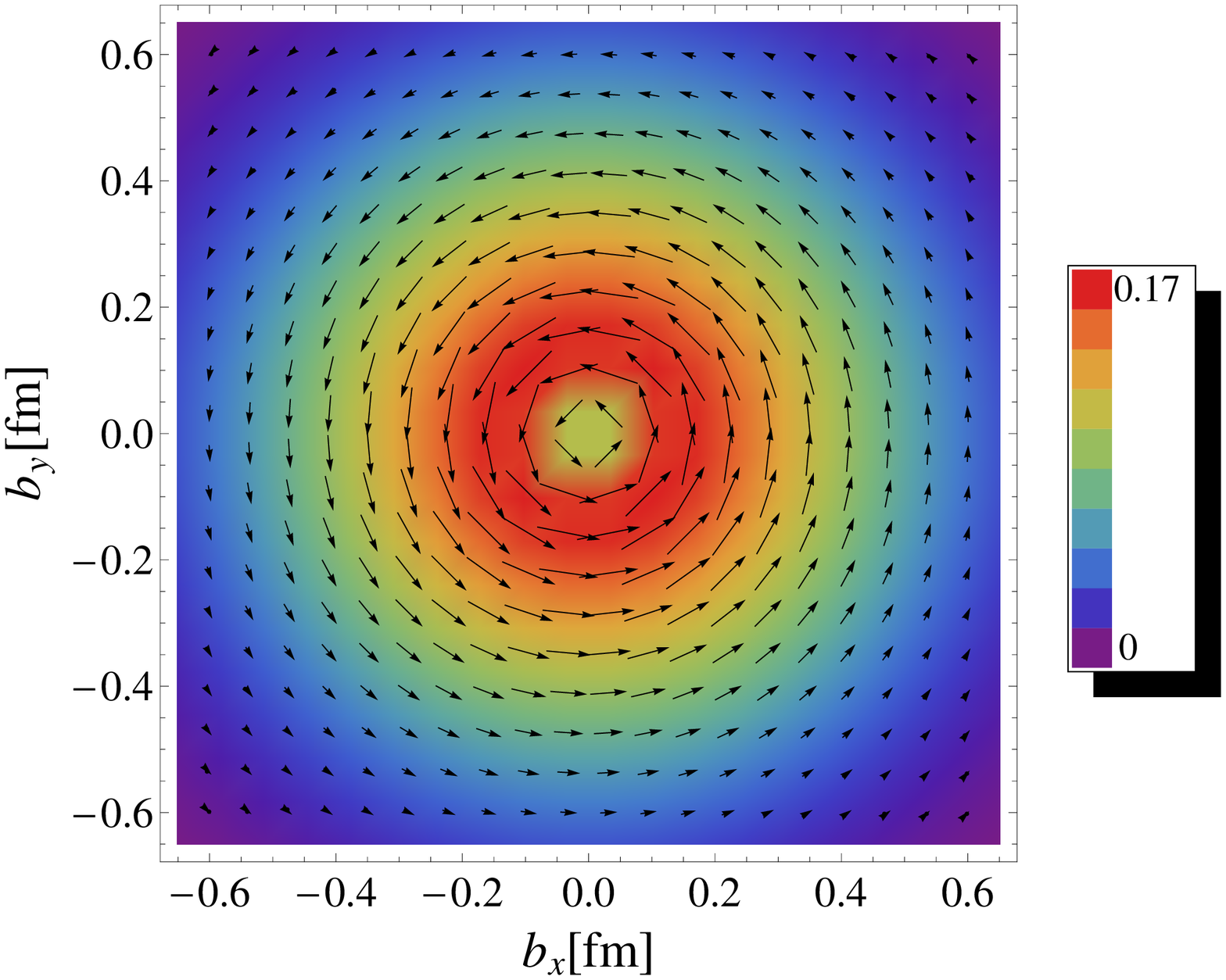}\hspace{1cm}
		\includegraphics[width=.38\textwidth]{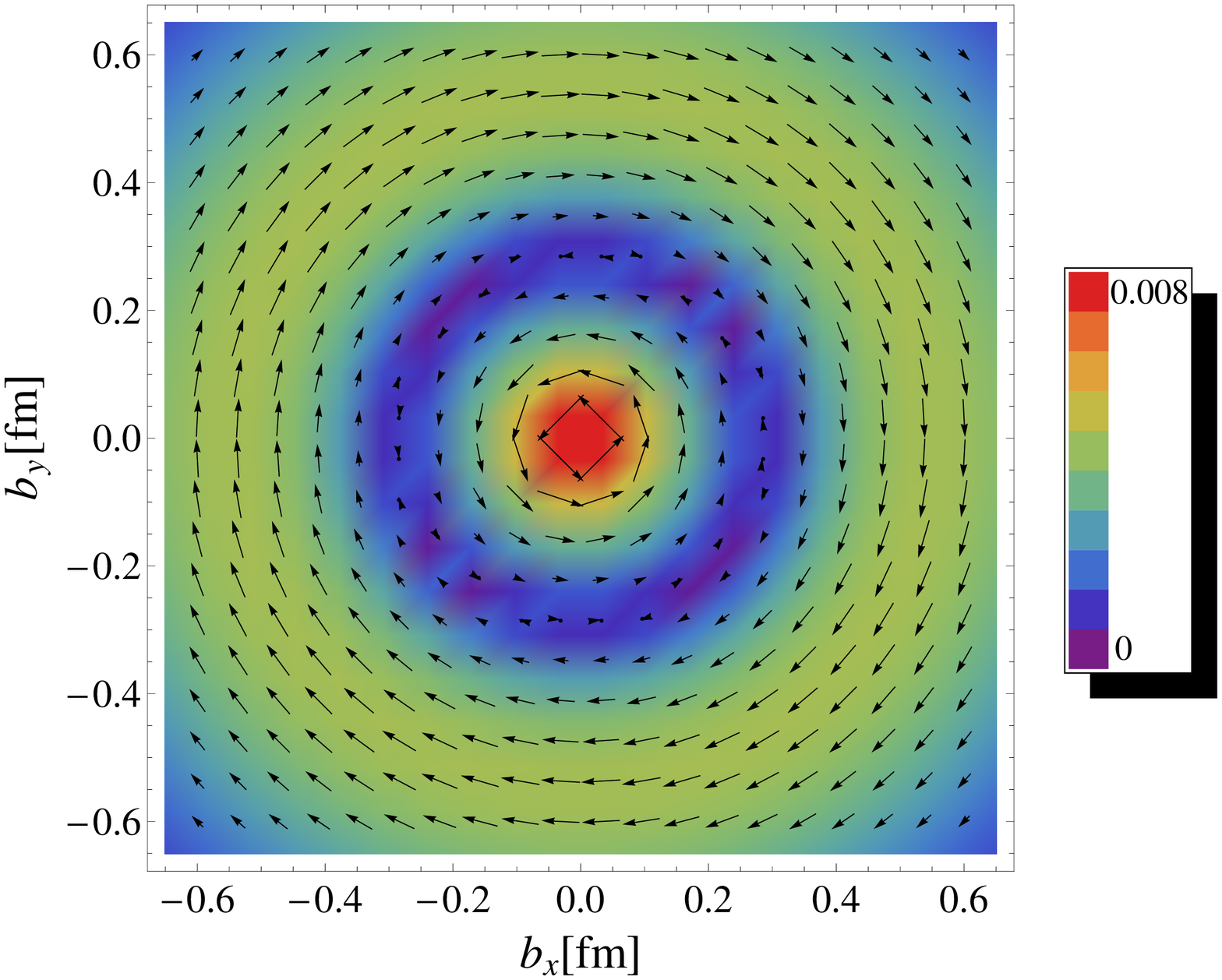}
	\caption{\footnotesize{The $x$-integrated distributions in impact-parameter space for unpolarized quarks in a longitudinally polarized proton (the proton spin points out of the plane). The upper panels show the distortion of the Wigner distribution, for a given transverse momentum $\vec k_\perp=k_\perp\,\vec e_y$ with $k_\perp=0.3$ GeV, induced by the proton polarization, and the lower panels show the distribution of the average quark transverse momentum. The left panels are for $u$ quarks and the right panels for $d$ quarks. These distributions have been obtained from the light-front constituent quark model~\cite{Lorce:2011kd}.}}\label{fig1}
\end{figure}

In fig.~\ref{fig1} we show the results in impact-parameter space obtained in the light-front constituent quark model. The upper panels show the distortions of the Wigner distribution for $u$ (left panels) and $d$ (right panels) quarks for a given transverse momentum $\vec k_\perp=k_\perp\,\vec e_y$ with $k_\perp=0.3$ GeV. In particular, the dipole structure indicates that $u$ quarks (resp. $d$ quarks) with OAM parallel (resp. antiparallel) to the proton polarization are favored. This can be even better seen in the lower panels of fig.~\ref{fig1} showing the distribution in impact parameter of the average quark transverse momentum $\langle\vec k_\perp\rangle(\vec b_\perp)=\int\ud x\,\ud^2k_\perp\,\vec k_\perp\,\rho^{[\gamma^+]}(\vec b_\perp,\vec k_\perp,x,\vec e_z)$. Interestingly, we observe that the $d$-quark OAM changes sign around $0.25$ fm away from the transverse center of momentum. 

\section{Quark orbital angular momentum}

The Wigner distributions are rather intuitive objects as they correspond to phase-space distributions in a semiclassical picture. In particular, any matrix element of a quark operator can be rewritten as a phase-space integral of the corresponding classical quantity weighted by the Wigner distribution. It is therefore natural to define the quark OAM as follows~\cite{Lorce:2011kd}
\begin{equation}\label{OAMWigner}
l_z^q=\int\ud x\,\ud^2k_\perp\,\ud^2b_\perp\left(\vec b_\perp\times\vec k_\perp\right)_z\,\rho^{[\gamma^+]q}(\vec b_\perp,\vec k_\perp,x,+\vec e_z).
\end{equation}
Since the Wigner distribution involves in its definition a gauge link, it inherits a path dependence. The simplest choice is a straight gauge link. In this case, Eq.~\eqref{OAMWigner} gives the kinetic OAM $L^q_z=l^{q,\text{straight}}_z$ associated with the quark OAM operator appearing in the Ji decomposition~\cite{Ji:1996ek,Ji:2012sj} $-\frac{i}{2}\int\ud^3r\,\overline\psi^q\gamma^+\left(\vec r\times\!\stackrel{\leftrightarrow}{D}_r\right)_z\psi^q$. According to the Ji's sum rule~\cite{Ji:1996ek}, this kinetic quark OAM can be extracted from the GPDs
\begin{equation}\label{ji-sumrule}
L^q_z=\frac{1}{2}\int^1_{-1}\ud x\left\{x\left[H^q(x,0,0)+E^q(x,0,0)\right]-\tilde H^q(x,0,0)\right\}.
\end{equation}

In order to connect the Wigner distributions to the TMDs, it is more natural to consider instead a staple-like gauge link consisting of two longitudinal straight lines connected at $x^-=\pm\infty$ by a transverse straight line~\cite{Meissner:2009ww,Lorce:2011dv}. In this case, Eq.~\eqref{OAMWigner} gives the canonical OAM $\ell_z=l^{q,\text{staple}}_z$ associated with the quark OAM operator appearing in the Jaffe-Manohar decomposition in the $A^+=0$ gauge $-\frac{i}{2}\int\ud^3r\,\overline\psi^q\gamma^+\left(\vec r\times\!\stackrel{\leftrightarrow}{\nabla}_r\right)_z\psi^q$~\cite{Jaffe:1989jz}. Recently, it has been suggested, on the basis of some quark-model calculations, that the TMD $h_{1T}^\perp$ may also be related to the quark OAM~\cite{Lorce':2011kn}
\begin{equation}
\mathcal L_z^q=-\int\ud x\,\ud^2k_\perp\,\frac{k_\perp^2}{2M^2}\,h_{1T}^{\perp q}(x,k^2_\perp).
\end{equation}
However, no rigorous expression for the OAM in terms of the TMDs is known so far. For a more detailed discussion on the different decompositions and corresponding OAM, see ref.~\cite{Lorce:2012rr}.

In Table \ref{OAMtable}, we present the results from the light-front constituent quark model (LFCQM) and the light-front version of the chiral quark-soliton model (LF$\chi$QSM) restricted to the three-quark sector~\cite{Lorce:2011kd}. 
\begin{table}[th!]
\begin{center}
\caption{\footnotesize{Comparison between the Ji ($L^q_z$), Jaffe-Manohar ($\ell^q_z$) and TMD ($\mathcal L^q_Z$) OAM in the LFCQM and the LF$\chi$QSM for $u$-, $d$- and total ($u+d$) quark contributions.}}\label{OAMtable}
\begin{tabular}{@{\quad}c@{\quad}|@{\quad}c@{\quad}c@{\quad}c@{\quad}|@{\quad}c@{\quad}c@{\quad}c@{\quad}}\hline
Model&\multicolumn{3}{c@{\quad}|@{\quad}}{LFCQM}&\multicolumn{3}{c@{\quad}}{LF$\chi$QSM}\\
$q$&$u$&$d$&Total&$u$&$d$&Total\\
\hline
$L^q_z$&$0.071$&$~~0.055$&$0.126$&$-0.008$&$~~0.077$&$0.069$\\
$\ell^q_z$&$0.131$&$-0.005$&$0.126$&$~~0.073$&$-0.004$&$0.069$\\
$\mathcal L^q_z$&$0.169$&$-0.042$&$0.126$&$~~0.093$&$-0.023$&$0.069$\\
\hline
\end{tabular}
\end{center}
\end{table}
As expected in a pure quark model, all the definitions give the same value for the total quark OAM, with nearly twice more net quark OAM in the LFCQM than in the LF$\chi$QSM. The difference between the various definitions appears in the separate quark-flavour contributions. Note in particular that unlike the LFCQM, the LF$\chi$QSM predicts a negative sign for the $u$-quark OAM in agreement with lattice calculations~\cite{Hagler:2007xi}. It is surprising that $\ell^q_z\neq L^q_z$ since it is generally believed that the Jaffe-Manohar and Ji's OAM should coincide in absence of gauge degrees of freedom. Note that a similar observation has also been made in the instant-form version of the $\chi$QSM~\cite{Wakamatsu:2005vk}.

\section{Conclusion}

In summary, we presented the first model calculation of the Wigner distribution free of relativistic corrections and discussed its connection with the quark orbital angular momentum. Using a light-front constituent quark model and the light-front version of the chiral quark-soliton model, we compared the various definitions for the quark orbital angular momentum.

\acknowledgments
C. Lorc\'e is thankful to INFN and the Department of Physics of the University of Pavia for their hospitality. This work was supported in part by the Research Infrastructure Integrating Activity ``Study of Strongly Interacting Matter'' (acronym HadronPhysic3, Grant Agreement n. 283286) under the Seventh Framework Programme of the European Community, by the Italian MIUR through the PRIN 2008EKLACK ``Structure of the nucleon: transverse momentum, transverse spin and orbital angular momentum'', and by the P2I (``Physique des deux Infinis'') project.


\begin{thebibliography}{99}

\bibitem{Ji:2003ak}
  X.~Ji, {\it Viewing the proton through `color'-filters}, Phys.\ Rev.\ Lett.\  {\bf 91}, 062001 (2003) [arXiv:hep-ph/0304037];\\
  A.~V.~Belitsky, X. Ji and F.~Yuan, {\it Quark imaging in the proton via quantum phase-space distributions}, Phys.\ Rev.\  D {\bf 69}, 074014 (2004) [arXiv:hep-ph/0307383].

\bibitem{Meissner:2009ww}
  S.~Meissner, A.~Metz and M.~Schlegel, {\it Generalized parton correlation functions for a spin-1/2 hadron}, JHEP {\bf 0908}, 056 (2009) [arXiv:0906.5323 [hep-ph]].

\bibitem{Lorce:2011kd}
  C.~Lorc\'e and B.~Pasquini, {\it Quark Wigner Distributions and Orbital Angular Momentum}, Phys. Rev. D {\bf 84}, 014015 (2011) [arXiv:1106.0139 [hep-ph]].

\bibitem{Lorce:2011dv}
  C.~Lorc\'e, B.~Pasquini and M.~Vanderhaeghen, {\it Unified framework for generalized and transverse-momentum dependent parton distributions within a 3Q light-cone picture of the nucleon}, JHEP {\bf 1105}, 041 (2011) [arXiv:1102.4704 [hep-ph]].
  
\bibitem{Ji:1996ek}
  X.~D.~Ji, {\it Gauge invariant decomposition of nucleon spin and its spin - off}, Phys.\ Rev.\ Lett.\  {\bf 78}, 610 (1997) [arXiv:hep-ph/9603249].

\bibitem{Ji:2012sj} 
  X.~Ji, X.~Xiong and F.~Yuan, {\it Seeking Partonic Pictures of Proton Spin}, arXiv:1202.2843 [hep-ph].

\bibitem{Jaffe:1989jz} 
  R.~L.~Jaffe and A.~Manohar, {\it The G(1) Problem: Fact and Fantasy on the Spin of the Proton}, Nucl.\ Phys.\ B {\bf 337}, 509 (1990);\\
  C.~Lorc\'e, B.~Pasquini, X.~Xiong and F.~Yuan, {\it The quark orbital angular momentum from Wigner distributions and light-cone wave functions}, Phys.\ Rev.\  D {\bf 85}, 114006 (2012) [arXiv:1111.4827 [hep-ph]];\\
  Y.~Hatta, {\it Notes on the orbital angular momentum of quarks in the nucleon}, Phys.\ Lett.\ B {\bf 708}, 186 (2012) [arXiv:1111.3547 [hep-ph]].

\bibitem{Lorce':2011kn} 
  C.~Lorc\'e and B.~Pasquini, {\it Pretzelosity TMD and Quark Orbital Angular Momentum}, Phys.\ Lett.\ B {\bf 710}, 486 (2012) [arXiv:1111.6069 [hep-ph]].

\bibitem{Lorce:2012rr} 
  C.~Lorc\'e, {\it Geometrical approach to the proton spin decomposition}, arXiv:1205.6483 [hep-ph].

\bibitem{Hagler:2007xi}
  Ph.~H\"agler {\it et al.}  [LHPC Collaborations], {\it Nucleon Generalized Parton Distributions from Full Lattice QCD}, Phys.\ Rev.\  D {\bf 77}, 094502 (2008) [arXiv:0705.4295 [hep-lat]].

\bibitem{Wakamatsu:2005vk}
  M.~Wakamatsu, H.~Tsujimoto, {\it The Generalized parton distribution functions and the nucleon spin sum rules in the chiral quark soliton model}, Phys.\ Rev.\  D{\bf 71}, 074001 (2005) [hep-ph/0502030];\\
  M.~Wakamatsu, {\it The Role of orbital angular momentum in the proton spin}, Eur.\ Phys.\ J.\  A{\bf 44}, 297 (2010) [arXiv:0908.0972 [hep-ph]].

  
\end{thebibliography}
\end{document}